**RESEARCH ARTICLE**

# Know Your Customer: Balancing innovation and regulation for financial inclusion

Karen Elliott[1] , Kovila Coopamootoo[2], Edward Curran[3], Paul Ezhilchelvan[4], Samantha Finnigan[5], Dave Horsfall[5], Zhichao Ma[4], Magdalene Ng[6], Tasos Spiliotopoulos[1] , Han Wu[4] and Aad van Moorsel[7],*

[1] Newcastle University Business School, Newcastle University, Newcastle upon Tyne NE1 7RU, United Kingdom
[2] Department of Informatics, King's College London, London WC2R 2LS, United Kingdom
[3] Atom Bank, Durham DH1 5TS, United Kingdom
[4] School of Computing, Newcastle University, Newcastle upon Tyne NE1 7RU, United Kingdom
[5] Digital Institute, Newcastle University, Newcastle upon Tyne NE1 7RU, United Kingdom
[6] Social Sciences, Westminster University, London W1W 6XH United Kingdom
[7] School of Computer Science, University of Birmingham, Birmingham B15 2TT, United Kingdom
*Corresponding author. E-mail: a.vanmoorsel@bham.ac.uk

**Received:** 17 December 2021; **Revised:** 29 March 2022; **Accepted:** 22 August 2022

**Key words:** decentralized identifier; digital identity; financial inclusion; know your customer; selective disclosure; self-sovereign identity; verifiable credentials

**Abbreviation:** DID, decentralized identifiers; FCA, UK Financial Conduct Authority; GDPR, general data protection regulation; KYC, know your customer; NHS, National Health Service; VC, verifiable credentials; W3C, world-wide web consortium

**Abstract**

Financial inclusion depends on providing adjusted services for citizens with disclosed vulnerabilities. At the same time, the financial industry needs to adhere to a strict regulatory framework, which is often in conflict with the desire for inclusive, adaptive, and privacy-preserving services. In this article we study how this tension impacts the deployment of privacy-sensitive technologies aimed at financial inclusion. We conduct a qualitative study with banking experts to understand their perspectives on service development for financial inclusion. We build and demonstrate a prototype solution based on open source decentralized identifiers and verifiable credentials software and report on feedback from the banking experts on this system. The technology is promising thanks to its selective disclosure of vulnerabilities to the full control of the individual. This supports GDPR requirements, but at the same time, there is a clear tension between introducing these technologies and fulfilling other regulatory requirements, particularly with respect to "Know Your Customer." We consider the policy implications stemming from these tensions and provide guidelines for the further design of related technologies.

**Policy Significance Statement**

The work presented in this article supports policymakers in two distinct ways. First, we inform policies that aim to support financially vulnerable populations and tackle financial exclusion. Drawing from specific guidance by the UK Financial Conduct Authority, we demonstrate how digital identity technologies can enable vulnerable consumers to disclose their vulnerability status with financial firms in a privacy-preserving way and we discuss how financial firms can use these data efficiently to comply with regulatory guidance and improve support for vulnerable populations, while ensuring that Know Your Customer requirements are being met. As digital identity technologies evolve and become more widespread, financial policymakers will benefit from considering these insights when creating new policies and adapting existing policies. In addition, by developing a working prototype and discussing its features and characteristics with experts in the financial domain, we inform emerging policies that focus on the deployment and support of digital identity technologies, in general, and decentralized identifiers (DIDs) and verifiable credentials (VCs), in particular, such as the *Digital Identity and Attributes Trust Framework* in the UK and similar nascent policy efforts worldwide.

**1. Introduction**

An important enabler for financial inclusion is the ability of citizens to disclose the vulnerability, to allow a bank or other financial service to train staff, allocate resources, develop products and design services in a way that supports vulnerable populations. In the UK, the Financial Conduct Authority (FCA) has been tracking the vulnerability of consumers over time, together with other aspects of their financial lives. FCA's flagship consumer survey, the Financial Lives survey, has found that, before COVID-19, the number of UK adults showing one or more characteristics of vulnerability was decreasing, with this decrease largely attributed to improvements in digital inclusion and financial resilience. However, the latest results of this survey show that COVID-19 has reversed this positive trend in vulnerability and has disproportionately affected specific population groups, such as younger adults and the self-employed (Financial Conduct Authority UK, 2021a).

Our research in financial inclusion has been heavily influenced by the FCA's latest report which provides guidance for firms on the fair treatment of vulnerable customers in finance (Financial Conduct Authority UK, 2021b). This guidance identifies a vulnerable customer as "someone who, due to their personal circumstances, is especially susceptible to harm—particularly when a firm is not acting with appropriate levels of care." The guidance establishes the protection of vulnerable customers as a key priority for the industry and strongly encourages financial firms to treat vulnerable customers with ample consideration of equity of treatment.

Exploring the four categories of characteristics that drive financial vulnerability—poor health, impact of life events, low resilience, and low capability, we investigate the use of Verifiable Credentials (VCs) and Decentralized Identifiers (DIDs) for identifying and supporting financially vulnerable consumers. There are two important characteristics that make DIDs and VCs desirable candidate technologies for financial inclusion solutions, namely self-sovereignty and selective disclosure. Self-sovereignty says that individuals control their own data, be it on their or other institutions' devices, providing these data only when someone needs to validate them. Selective disclosure says that only directly relevant private information is shared with interested parties in a privacy-preserving way. Spiliotopoulos et al. (2021) explain this further and examine in detail an overarching scenario that involves a customer interacting with their bank using VCs for different purposes and in different ways (e.g., both directly and via a revocable token), as well as the possible response and use of the vulnerability information by the bank.

We engage with representatives of the financial services sector both in advance and after completion of the system implementation. This uses a qualitative approach, conducting in-depth interviews with a selective group of experts, and we refer to Section 5.1 for further discussion of the qualitative research methods used in this work. This qualitative approach provides us with a unique up-to-date view financial services experts have on the topic of financial inclusion, which is arguably increasingly important to their sector given various ethical and societal pressures.

This article is organized as follows. In Section 2, we first discuss with experts how the financial industry currently addresses financial inclusion and vulnerability disclosure. Section 3 then provides an overview of technological approaches to the disclosure of credentials, usually integrated with identity systems. We built a working prototype of vulnerability disclosure using emerging DID and VC technology, on top of the Microsoft Identity Overlay Network, see Section 4. This prototype was presented to the group of banking experts and Section 5 provides their responses to the DID- and VC-based approaches. This section also explains the qualitative research methods used (Section 5.1). We conclude the article with suggestions for policy makers and regulators in Section 6.

## 2. Vulnerability Disclosure: Current Approaches in the Financial Services Sector

To understand how financial institutions have approached vulnerability in light of the FCA guidance (Financial Conduct Authority UK, 2021b), we accessed stakeholders from the mainstream financial services sector (e.g., senior financial support agents). In addition, we spoke to the alternative responsible lending sector, such as directors of credit unions and managers of financial community interest groups providing support in navigating financial systems. All stakeholders have over 20 years experience of working in the financial services sector, within the alternative provider cohort, mainstream knowledge was transferred in assisting customers emanating from disadvantaged and vulnerable cohorts which permits this study to understand this space from a business perspective (mainstream and alternate) whilst capturing reported customer experience (See Table 1 in Section 5.1 for detail on the consulted stakeholders.).

We asked the mainstream stakeholders how their organizations perceive vulnerability and explore current procedures to identify the customer group. Mainstream institutions reported that current processes were predominantly manual, based on providing vulnerability identification and support discerned from customer phone conversations and chat messages with financial agents. During this interaction, vulnerability disclosure was raised by the customer rather than the agent and this invoked a response and categorization by the mainstream provider. An example of vulnerability that emerged was revealed as encompassing a range of factors from a lack of digital illiteracy to physical and sensitive conditions impacting an individual's financial capabilities and resilience in dealing with their financial responsibilities. Interestingly, customers willingly disclose physical conditions, such as blindness or deafness, as opposed to more sensitive conditions, surrounding mental health issues, and mainstream providers believed that this reticence was premised on associated social stigma (Dewa, 2014; Zolkefli, 2021). However, for terminal illness or bereavement, customers again are reported as open and engaged in voluntary disclosure.

Once vulnerability was confirmed by the mainstream provider, what happened to the customer's accounts? We found that the term vulnerable is replaced with "additional customer care and support" protocols to avoid negative labeling of customers. In addition, "flagging" policies are then attached to customer records with a review system recurring every 12 months to ensure applicability. To flag an account, customers and agents liaise to agree on the nature of the vulnerability, how these impact on their financial capacity, the support required, and how customer data is stored and used by the mainstream provider (UK GDPR compliance, 2018). Flags are read by agents each time the customer accesses services with flexibility to assign an account manager if account arrears accrued and monitored every 3 months. If the issue is resolved the flag is removed. The customer can also request flag removal, and information is no longer stored (cf. GDPR, *the right to be forgotten*). Thus, stakeholders purport to deliver flexible and tailored customer service for vulnerabilities.

*Table 1.* Financial inclusion experts (N = 5)

| Gender | Age | Education level | Tenure in finance | Current role |
|---|---|---|---|---|
| Female1 | 50+ | MLIA (Dip) Finance | +40 years | Director of Credit Union |
| Male1 | 50+ | HND Business | +20 years | Independent mainstream and alternative financial consultant |
| Male2 | 40+ | Bachelor's degree | +15–20 years | Independent mainstream and alternative director and founder of financial provider |
| Male3 | 40+ | Post-graduate degree | +20 years | Manager/Founder of community financial group for disadvantaged communities |
| Female2 | 50+ | Post-graduate degree | +20 years | Founder of social and financial inclusion group collaborating with mainstream and alternative providers |

We enquired how the engagement process commences with customers in establishing vulnerability. As discussed, phone conversations between the mainstream and alternative providers and customer are the current primary route however, new functionality is being explored for sensitive conditions. For instance, as we noted that disclosure of mental health conditions is compounded by perceptions of social stigma, automatic notifications via the Vulnerability Registration Service[1] have recently been added to institution's protocols to inform mainstream and alternative providers of a vulnerability disclosure that is registered by the customer and provides consent to access the details for the financial provider.

To compare current processes with verifiable credential technology, we discussed emerging solutions for disclosure: selective disclosure on external cards, or a dedicated vulnerability card. The mainstream response was negatively premised on part of the design as a binary disclosure card that is, "are you vulnerable—yes or no" which was deemed inappropriate as with insufficient data, mainstream providers claimed to be liable to operational risk. Specifically, expressing concern that the current requirements from the FCA regulatory body *know your customer* (KYC), may not be fulfilled in the use of verifiable credentials, and thus, risk of fines from the FCA may occur which could tarnish mainstream provider's reputation in the sector. Indeed, selective (attribute) disclosure was viewed as aligned to suspicious frequent patterns of fraudulent behavior. Yet, the experts we consulted suggest that at most half of customers who are vulnerable are captured using the reported manual process, meaning many are left to reside in a vulnerable status despite existing support mechanisms.

For alternative stakeholders (i.e., responsible lenders, credit unions, etc.), algorithmic tools are available to onboard customers ensuring that responsible lending is observed, in response to the use of verifiable credential technology in identifying and disclosing vulnerability, this cohort cited technological and behavioral challenges. From a technology perspective, the question was raised whether decentralized technology can or should share profiles with third parties when the transaction data is used for credit decision-making processes. An *aggregation hub* to consolidate the digital profiles and make them accessible for decisioning was suggested. The binary nature of disclosure was also questioned in ensuring how potential different verifiers of credentials demonstrate digital transaction histories for vulnerable customers.

From a behavioral perspective, the question was raised by both sets of stakeholders on how to motivate customer engagement in tracking their financial behavior, moving to verifiable technologies, and how to communicate the benefits of use, especially, as citizens may trust the anonymity of cash transactions. The adoption of M-PESA in Africa serves as an example of the advantages of credentials, whereby technology verifies the customer identity while granting similar anonymity to cash during transactions and a proven customer credit history is unnecessary (Burns, 2018; Van Hove and Dubus, 2019). In terms of disadvantages, mainstream decision-makers, raised professional behavioral questions, concerning the risk of collusion and fraud between verifiers in the credential process, for instance, materializing in creating fake profiles and what safeguards existed to prevent a proliferation of profiles from emerging. To explain, the credentials process includes three actors, the issuer (as we use later the National Health Service), the verifier (financial institutions and lenders), and the holder (customer) premised on the immutable properties in using technologies (for full details, see following sections). Hence, if collusion occurs between verifiers, this may lead to the detriment of institutions to unscrupulous competitors defrauding the credential process. In short, undermining the advantages of disclosing vulnerability for financing and customer support to drive inclusion whilst breaking trust perceptions for all parties involved in the credential process. Furthermore, an overarching issue for institutions, involved regulatory obligations and how any credential solution fulfils KYC obligations from the FCA, if institutions could not verify, differentiate or sufficiently validate the credential. Therefore, leading to the risk of irresponsible lending on behalf of the financial institution.

---

[1] https://www.vulnerabilityregistrationservice.co.uk/.

## 3. Selective Sharing of Vulnerabilities Using Verifiable Credentials

To disclose vulnerabilities, identity systems need to be augmented with the ability to associate certain characteristics (also called claims, attributes or credentials) to an identifier. Identity systems have been surveyed extensively, for instance recently in Beduschi (2021), and in this article we therefore particularly consider the role of credentials within such systems.

### 3.1. Definitions and concepts

Cameron (2005) defines digital identity to be "a set of claims made by one digital subject (e.g., a user) about itself or about another digital subject." These claims are asserted truths of a subject, that is, attributes, including biometric information (e.g., fingerprints), life factors (e.g., data of birth), and qualifications (e.g., degree certificates). A digital identity system is required to reliably deliver identifying information of one subject to another subject while detecting deception. This process introduces the concept of identity verification, which confirms and establishes a link between a claimed identity and the actual, living person presenting the evidence (Beduschi, 2021).

Self-sovereign identity is an emerging type of digital identity which has been widely studied but loosely defined, for example, Mühle et al. (2018). In the influential work of Allen (2021), self-sovereign identity is defined to be transportable, which means it cannot be locked down to one site or locale, and self-sovereign identity must also allow ordinary users to make claims, which could include personally identifying information or facts about personal capability or group membership. It can even contain information about the user that was asserted by other persons or groups. Sovrin Foundation is a non-profit global consortium aiming towards building and governing a network of self-sovereign identity (Tobin and Reed, 2016). Their definition of self-sovereign identity highlights three crucial properties: individual control, security, and full portability. They also state that "claims made about the user in identity transactions can be self-asserted, or asserted by a third party whose authenticity can be independently verified by a relying party."

Decentralized Identifiers are a type of identifier that enables a verifiable, decentralized digital identity, and are based on the self-sovereign identity paradigm of the W3C standardization consortium (World Wide Web Consortium (W3C), 2021). Verifiable credentials, formerly known as verifiable claims, is defined as the union of different assertions about a self-sovereign identity. Here an assertion is a signed claim bound to an entity. It can be signed by the entity itself, creating the notion of a self-asserted claim. Alternatively, it can be signed by the provider, acting as the trusted third party, creating a provider-asserted claim.

### 3.2. Multiple stakeholders in verifiable credentials

In VC approaches such as World Wide Web Consortium (W3C) (2021), DID-based URLs are used for expressing identifiers associated with *subjects*, *issuers*, *holders* and other machine-readable information associated with a verifiable credential, see Figure 1. Note that in principle verifiable credentials are not dependent on DIDs and DIDs do not depend on verifiable credentials. However, many verifiable credentials implementations adopt DIDs, and software libraries implementing the specification resolve DIDs. The core actors and the relationships between them can be represented with a chain of trust, as in

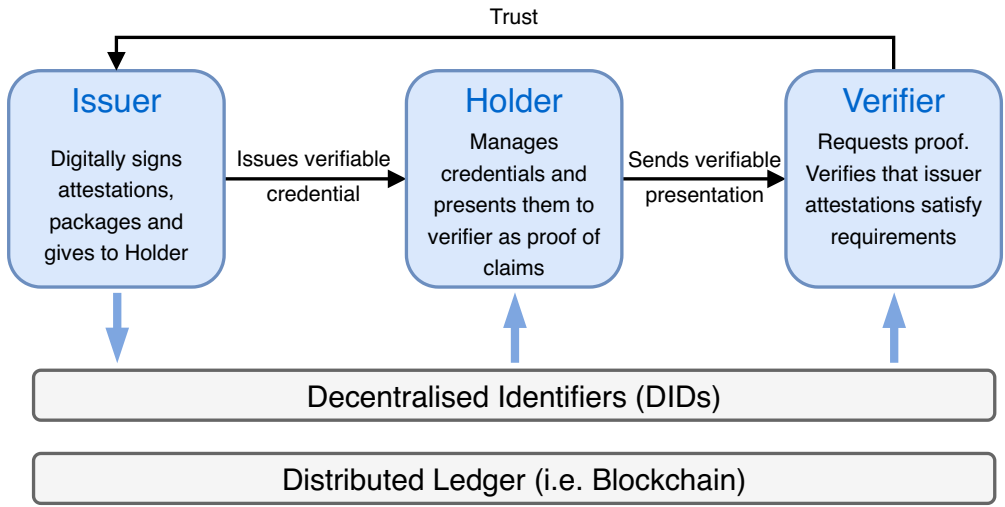

**Figure 1.** *The chain of trust, adapted from World Wide Web Consortium (W3C) (2021).*

Figure 1. Issuers create credentials, holders store them, and verifiers ask for proofs of claims based upon them.

All entities trust the verifiable data registry to be tamper-evident and to be a correct record of which claims are controlled, by which entities. A distributed ledger (i.e., Blockchain) is used to establish a chain of trust between stakeholders. The holder and verifier trust the issuer to issue legitimate credentials about the subject, and to revoke them quickly when appropriate. The holder trusts the repository to store credentials securely, to not release them to anyone other than the holder, and to not corrupt or lose them while they are in its care.

The concept of selective disclosure means that a derived verifiable credential is formatted according to the verifier's data schema instead of the issuer's data schema, without needing to involve the issuer after verifiable credential issuance. This provides a great deal of flexibility for holders to use their issued verifiable credentials, and improves privacy. The selective disclosure schema is enabled by Zero-Knowledge Proofs method, which allows holders to indirectly prove they hold claims from a VC without exposing the VC. For instance, a claim that contains a subject's date of birth can be used to confirm the subject's age is within a given range. This is sufficient for age-related qualifications, without revealing the actual birth date. The holder has the flexibility to use the claim in any way that is applicable to the desired verifiable presentation.

A number of security concerns should be highlighted for issuers, holders, and verifiers when processing data with verifiable credentials. First, the cryptography suites and libraries need to be upgraded to the newest version as they have a shelf life and can be vulnerable to new attacks. VCs usually contain website links that may direct users to data outside of VC (e.g., images, text), which are not protected by the proof on the VC. In this case, encrypting the website links is necessary to reduce such risk. While using the selective disclosure scheme, the issuer is required to bundle dependent claims securely in case that the holder bundle different credentials in a way that is not intended by the issuer.

### 3.3. Evolution of digital identity and credential models

The management of digital identities has always been regarded as a cornerstone for efficiently organizing resources and services in the Internet era. In the past decades, digital identity technology has evolved from *isolated* to *centralized*, then to *federated* and *user-centric*, and is now moving towards *decentralized* models (Zhu and Badr, 2018).

In the isolated model, service providers act as both credential provider and identifier provider by controlling the name space for a specific service domain, and allocating identifiers to users (Jøsang and Pope, 2005). This means that the user must manage as many identifiers and credentials (e.g., passwords) as the service providers that users transact with. The drawback of this model is apparent: the user has to memorize a large number of logins and passwords. Consequently, users may choose the same password for multiple accounts, creating a significant risk to user security levels, if any server(s) is attacked and the password is recovered.

The centralized model is designed to solve the problems mentioned above (Jøsang et al., 2007). The identifier provider in this model centralizes digital identity management, which allows several service providers to rely on the same identity provider. Users can use the same identity and credentials to authenticate themselves with multiple service providers simultaneously, without repeating this process for a new service provider. While the number of identities for each user is significantly reduced in the centralized model, access to distributed services managed by different centralized systems and security domains remains unsupported.

The federated model groups service providers together to form a federation of identities, enabling service providers to recognize user identifiers and entitlements from other service providers within the same federated domain (Maler and Reed, 2008). To establish trust relationships among different service providers in this federated domain, both commercial agreements and technology supports are required. Examples of federated identity platform technologies include Shibboleth (Morgan et al., 2004), open-source architecture OpenID Connect (Bodnar et al., 2016) and WS-Federation by Microsoft and IBM (Goodner et al., 2007).

While the above models indeed mitigate the complexity of managing numerous identities across different security domains, they are designed from the perspective of service providers. This means that the user experience can be poor when the number of online service providers is increasing (Alpár et al., 2013). Therefore, user-centric models have been proposed to allow users to take complete control over their personal attributes (Angin et al., 2010). For example, Jøsang and Pope (2005) propose a user-centric digital identity system that depends on personal trusted devices (e.g., smartphones) to manage users' identities from different domains. OpenID 2.0 is another example of user-centric digital identity system for web services (Recordon and Reed, 2006) and Suriadi et al. (2009) also follow user-centric design principles to take into account user privacy in identity systems. Ahn et al. (2009) enhance user privacy for user-centric identity systems by applying privacy labels to personal claims. However, in such systems, users must rely on third parties, the identifier providers, to access services from different domains, which means users' transactions are exposed to those identifier providers.

Blockchain technology has emerged as a critical enabler in addressing the above challenges in digital identity systems (Alharbi and Hussain, 2021). The decentralized trusted nature of blockchain allows users to ensure the privacy and security of their personal data without relying on third parties. Self-sovereign identity is enabled under this type of decentralized model, which grants users the right to selectively disclose their attributes (Toth and Anderson-Priddy, 2019). Allen (2021) proposes 10 principles of implementing self-sovereign identity and Tobin and Reed (2016) further groups these principles into three categories: *security*, *controllability*, and *portability*. Examples of such blockchain-based identity systems include uPort (Panait et al., 2020) which is based on Ethereum and Blockstack (Ali et al., 2016).

Verifiable credentials technology is typically integrated with decentralized digital identity systems. In 2019, World Wide Web Consortium (W3C) (2021) issued a formal recommendation of verifiable credentials as digital documents issued with digital signatures, which are protected from corruption by asymmetric (public/private key) cryptography. For enhanced privacy, zero-knowledge proofs can additionally be used to reveal only the minimum of information required in an interaction, which is called selective disclosure. For example, a VC holder can choose to disclose a VC only containing the claim of date of birth, which can be used to derive the presented value to be over the age of 18, in a cryptographically verifiable manner. No additional personal information, also not the precise date of birth, need to be shared. In the age of increasing digital interactions and analysis of user data, self-sovereign identity could become the next stage in the evolution of digital identity systems. We turn to illustrate a prototype utilizing verifiable credentials technology applied to declaring vulnerabilities in the financial services industry.

## 4. An Implementation of Verifiable Credentials System for Declaring Vulnerabilities

### 4.1. System overview

We implemented a verifiable credentials system for vulnerability disclosure based on the World Wide Web Consortium (W3C) standards for DIDs v1.0 and Verifiable Credentials Data Model 1.0, running in an Azure environment. We defined a data model for a new Verifiable Credential type that maps the aforementioned drivers (poor health, impact of life events, low resilience, and low capability) of financial vulnerability to attributes. This allows the presentation of tamper-evident claims that cryptographically prove who issued them, but without the need to disclose the specific details of the vulnerability about which the claim is made.

The new credential type that asserts claims about vulnerability criteria, called VulnerabilityStatusCredential, has been defined with a schema built by extending existing vocabulary already available on the web at schema.org. We conducted interviews with partners in the financial industry who would verify these credentials to understand how they intend to request and consume them. To ensure interoperability of this credential, feedback from these sessions has been used to refine the credential type, schemas, and URIs for future use in the financial industry.

We define user roles, and the relationship between them to establish a triangle of trust between the Holder, the Issuer, and the Verifier (as in Figure 1). This trust model of VCs differentiates itself from most other trust models by ensuring that the Issuer and the Verifier do not need to trust the repository, and that the Issuer does not need to know or trust the Verifier. Our software can issue Verifiable Credentials that attest information about users, who can then present the credential enabling claims to be verified.

Through the aforementioned engagement with stakeholders and extensive attention to consumer needs, we have defined user-centered workflows that vulnerable users may encounter when trying to access financial services. Specifically, we engaged with financial personnel with a specific focus on "customer support and care" and substantial experience dealing with vulnerable customers. We tried to understand their process for identifying vulnerabilities and vulnerable customers, the different ways that vulnerabilities and vulnerable customers are classified, and what happens after a customer is identified as vulnerable. These discussions highlighted a clear tension around the disclosure of vulnerabilities. On one hand, some customers can be very anxious about sharing sensitive information because they are worried that it will be used against them. In addition, on a personal level they may feel embarrassed and are concerned that there is a stigma associated with some vulnerabilities. A rough estimate was that fewer than half of vulnerable customers were captured with current processes. On the other hand, a financial institution is interested in capturing as much of the potential vulnerability information as possible in order to adhere to regulatory guidance, prevent fraud and ensure the provision of more appropriate and tailored services to customers.

These discussions informed the design of our prototype and provided further context for the evaluation interviews. In particular, these findings suggested that a technological solution built around self-sovereignty and selective disclosure could mitigate customer concerns, prevent repeated disclosures from customers, promote data minimization and reduce the potential for fraud.

The architecture leverages services from Microsoft that facilitate the creation, storage, and presentation of Verifiable Credentials on the Identity Overlay Network (ION), which is a public DID overlay network. Two associated Node.js applications have been developed using the Microsoft VC Software Development Kit that issue VCs to end users and verify VCs from end users. The user interface is shown in Figure 2 to provide a sense of how citizens would experience the use of verifiable credentials using a mobile phone.

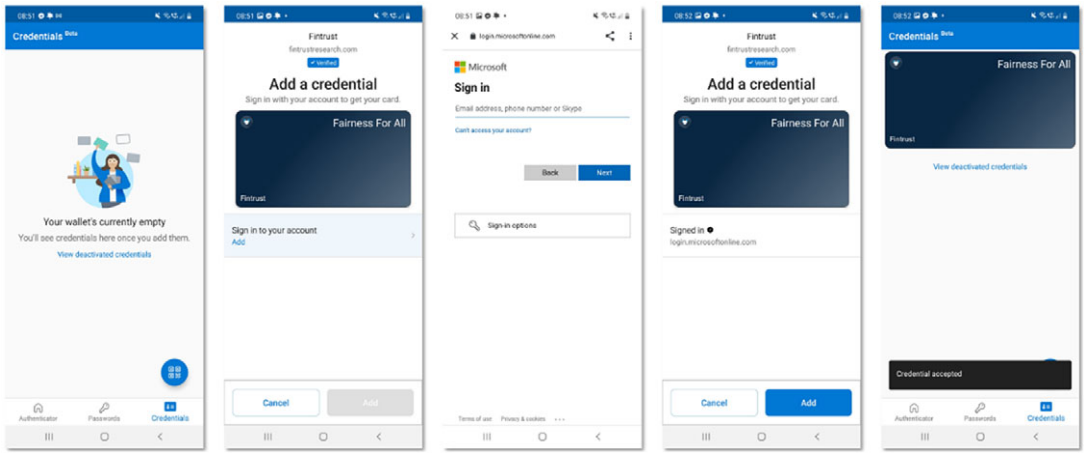

*Figure 2. User experience for Holder (cf. Figure 1) of credentials in the prototype implementation.*

## 5. Evaluation of Verifiable Credential Approaches to Vulnerability Disclosure

In this section, we use expert assessments to evaluate if verifiable credentials are suitable for the disclosure of vulnerabilities (Section 5.2). But first we introduce in Section 5.1 the research methods we used and the research questions we pursue.

### 5.1. Method and research questions

We had good access to stakeholders involved in technology, vulnerable customer engagement and financial inclusion aspects of the finance industry, such access is often a major barrier for research. The process of enquiry for this study was two-fold, first, during the design phase of the prototype described in the prior sections, we liaised with mainstream providers to understand the current vulnerability disclosure process with senior financial support agents ($n = 2$). Second, for the evaluation aspect of the study, we developed the prototype informed in part by the agents testimony and VC research (above), whilst our primary evaluation method was semi-structured interviews to understand concerns surrounding *fit-for-purpose* adoption in addressing vulnerability disclosure with mainstream and alternative providers. We used a cross-section of representative participants in the financial industry, see Table 1. Given the short duration of the project (10 months), we focused on a purposive sample of five experts (Etikan et al., 2016; Campbell et al., 2020), details of their demographic information are displayed in Table 1 below, maintaining anonymity, as suggested by Denis et al. (2001). The interviews commenced by requesting participants to describe their experience within the financial industry. Next, presented the FCAs four characteristics driving vulnerability and walked participants through the verification architecture and process from the individual perspective, culminating in Figure 2, followed by a series of questions to examine the feasibility and ease of use for both providers and customers (see Section 4.1). Because of ongoing COVID-19 restrictions, the interviews were conducted via Zoom using the *live transcript* function between May to June 2021, information and consent forms were distributed, signed, and returned before the interviews taking place (see Appendix I in Elliott et al., 2021). Each interview lasted for approximately 1 hr with some variation, following the protocol designed by the team (see Appendix II in Elliott et al., 2021).

A central part of the interview was to describe how the prototype would function as depicted in Figure 2. In short, the National Health Service (NHS) were selected as the use case for vulnerability in issuing a credential involving use of an individual's NHS number (identifier), as opposed to the usual three forms of identity, passport, drivers license, and a utility bill (for KYC). We tried to think of alternative credentials as many vulnerable customers do not possess these identification documents preventing financial inclusion. This NHS credential (among other information) holds some attributes (date of birth, test date, disability, etc.) and such attributes can be used to make claims about the holder (e.g., age > 18). The verifier (i.e., a financial provider) makes a request in order to verify a claim (i.e., enough information for a specific claim). We explained the caveat, that our implementation based on Microsoft ION technology currently only permits the sharing of a full credential (i.e., full NHS record, name, date or birth and address dependent on the credential issued) whereas, selective disclosure (see position article) posits that only an attribute (or a predicate on an attribute) is required to verify a claim, thus preserving the privacy of the user (a future development in ION). Our subsequent questions focused on evaluating the prototype from the expert's perspective to address our key research questions:

- *RQ1*. How can we promote user adoption of DID and VC technologies in the financial sector?
- *RQ2*. How can we maximize user disclosure of information in a privacy-preserving manner using VCs?
- *RQ3*. How can financial firms use DID/VC technologies efficiently to improve support for vulnerable populations?

Post-interview, the recordings were transcribed and to ensure participants retain maximum control over their information, transcripts were triangulated via verification with interviewees (Gioia et al., 2013). We have anonymized parts of the interviews included in this section, to ensure that specific individuals cannot be identified from the data presented. All interviews were conducted by the co-investigator of the research team.

Following Strauss (1987) and Corbin and Strauss (2015), we analyzed the subsequent data using a "process" coding approach utilizing NVivo software (v.12). This is a cyclical approach, where the general meaning of the discussions conducted within the semi-structured interviews is initially categorized (initiation), structured around specific themes (focus) and then reviewed and encoded (axial coding). All the material was reviewed to check that we had grasped what was significant to the interviewee (respondent validation; see Charmaz (2006)). Subsequently, items were reduced into a more manageable form of themes or "sets" (Gioia et al., 2013). To further enhance the validity of our findings, we include in the results and discussion extensive verbatim descriptions of the expert's views, to reduce the impact of our own biases.

### *5.2. Expert feedback*

This section presents the participant responses to the DID/VC presentation premised on the experts' experience within the financial services sector specifically, seeking to improve vulnerability support and promote financial inclusion using technological solutions. Participants were asked to consider the question from both a business and community/individual perspective.

#### *5.2.1. RQ1. How can we promote user adoption of DID and VC technologies in the financial sector?*

The challenge for the experts in terms of promoting adoption of the technologies in the financial sector is twofold. First, persuading mainstream and alternate stakeholders that technologies would bring return on investment in investing in the technologies, training staff, and so forth. Second, gaining the trust of customers to understand and use the application (e.g., a wallet function) to share credentials in the manner described in the prototype presentation. One expert expressed concern that stakeholders would be reluctant to promote the technologies unless the regulators provided a clear indication that such innovation was recommended (at present the FCA vulnerability report is guidance not compulsory):

> [I]t's also getting the FCA to acknowledge, approve, enhance their own requirements from a regulatory perspective for this, the adoption of this [technology]. I think that's absolutely key because as I say whether it's a fully regulated bank or a FinTech, who is sponsored by another third party, another issuer, how many of them are operating in this space? They will not move until the FCA has given its blessing and acknowledges and even promotes the applicationapplication… But, I think even to pilot something, such as this would be very, very hard to do without the FCA approval. (Male2)[2]

Both industry stakeholders and regulators are included in this description, thus, communication needs to be both internally to the stakeholders and to the broader general public by the FCA to facilitate and promote engagement with the technologies and associated benefits. The complexity added by these broader social actors therefore needs to be accounted for in relation to adoption via future collaboration with industry and regulators.

A further challenge was raised in terms of the financial stakeholders gathering information via such technologies and the customers being able to trust this party to ensure their data remained safe:

The banks have not served the wider community well from the financial crash. And the whole reason why we have got open banking is to create more competition for all…if they [banks] are given information about a user and do you remember when aids was a big thing…people wanted to keep HIV private…if they had to disclose it from for a mortgage and things like that, and people did not want to disclose it…with the credentials is, it depends what's in there…[t]rust is everything. And…fairness for all sounds trustworthy. I think it's a great title for it. (Female2)

The challenge is compounded by the impact of the global financial crisis (Pedersen, 2021) and recent instances where financial providers have breached regulatory rules around "know your customer" leading to more stringent examination of how providers analyze customer data (cf. Wirecard scandal BBC News, 2021). This has resulted in *"the issuers, financial institutions and FinTechs being particularly… very, very nervous and very risk averse"* (Male2) and mistrust by customers. An expert also raised concern over the cyber security aspect from the business perspective which could influence adoption of the technologies:

[H]ow do you ensure that I'm not a super-hot tech savvy guy? How do I or how does the bank know that I have not hacked into it [VC] and have just increased my benefits from 30 pounds a month to 5000? (Male2)

These extracts show two key issues facing financial services providers and customers' trust in adopting technologies. On the one hand, the perspective to be compliant with the FCA regulatory expectations suggests providers are prepared to reserve caution to meet all users' needs. On the other hand, in adopting technologies to garner more customer information specifically, around sensitive vulnerable details, skepticism exists surrounding the banks motives, vested interests (Alford, 1975) and general mistrust of the sector (Edelman, 2019). The use of experimental sandbox facilities allows for testing of alignment with FCA regulations.

*5.2.2. RQ2. How can we maximize user disclosure of information in a privacy-preserving manner using VCs?*
For financial services to provide care to vulnerable customers/users, the users will have to first, disclose the information pertaining to their vulnerability to the financial institution. However, as mentioned earlier, this sector suffers from mistrust amongst customer and the general public (Edelman, 2019). Therefore, we wanted the experts to consider how we could reassure customers that the technology can move towards the concept of self-sovereign identity (Der et al., 2017) permitting the customer to control the disclosure of vulnerabilities to the financial institution. As part of the protocol, we asked the experts to reflect on two scenarios regarding facilitating disclosure afforded by the technology.

---

[2] Throughout the article, we use brackets to denote text the authors added to the participant excerpts for clarity.

*5.2.3. Scenario 1*

Scenario 1 purported that a national vulnerability scheme, for example, the "Fairness for All" scheme is approved and initiated by the FCA. In short, issuers of a verifiable credential may include financial firms, other firms (e.g., the NHS, universities) and professional individuals (e.g., private doctors). For example, after a visit to the NHS, the NHS will issue the "normal" VC with all the information that would be included in an NHS certificate, as well as a "Fairness for All" VC that has limited information (attributes) and is only intended to be shared with financial institutions. Thus, the "Fairness for All" VC may say that the VC holder is not able to work for the next 2 months due to a disability but, will not disclose any further details about the disability. When the customer interacts with a bank, the bank simply asks for any "Fairness for All" VCs with a message like "The bank would like access to your credentials based on the Fairness for All scheme—Accept Y/N" and the customer can decide whether to present the VC to the bank.

*5.2.4. Scenario 2*

Scenario 2 concerned no public scheme exists as described above. Instead, when a customer interacts with a bank, a general mobile wallet application opens, and a request may appear as "The bank would like access to your verifiable credential repository for the following information: Are you $> = 18$ years? Do you have a physical disability? Do you have mental health problems? And so forth, Grant access Y/N?" (see Appendix II in Elliott et al., 2021). We sought the experts' opinion on how best to communicate such questions, tenets of the scheme, the process, and benefits in preserving vulnerable information to the customer/user groups. Furthermore, based on the discussions featured in Section 1, would consideration of a "vulnerability flag" or "score" be appropriate for both financial institution and customer vulnerability disclosure needs?

In response, one expert felt that current systems provided under the open banking system introduced over the past 5 years in the financial sector could assist in the demonstration of benefits of disclosure (Omarini, 2018):

> [W]hen you are talking about people being reluctant to identify themselves as…a vulnerability event. I was thinking…GDPR, there are different rules on the data and security. For example…we have basic, often very, very basic information on members, but if we started to ask questions such as and more personal questions around…whatever it is that is deemed to be sensitive data, then that increases data stored…I'm thinking maybe this is open banking at one level, and then maybe there's little offshoots of that, where almost, it's like an option, you can opt into providing the other information that's needed…saying this vulnerability factor? (Female1)

In addition, the linkage between the move towards open banking and eventually open life was highlighted in communicating to customers/users the benefits of disclosure to improve individual financial journeys (Sclove, 2020):

> A good comms plan around it [technology] to make it user friendly for everybody to understand. But it's pretty intuitive and I think that it could be adopted, I like the thinking about it from a comms point of view, fairness for all, because it's so important now that we have this democracy of life, but also, democracy of technology. (Female2)

Specifically, examining the technological use case and benefits, an expert espoused their perspective on VC technologies and the eventual progress to self-sovereign identities:

> There should not be any difference between someone that is vulnerable or disabled, as tech enables people to be augmented with more conforming to normality, because tech starts to take those disadvantages away…you can promote being vulnerable [using this technology] in a way that is now enabling better service and better support. (Male1)

Furthermore, stating that we should examine history to learn lessons from disability and vulnerability disclosure schemes without revealing the "full" credential, Male1 responded "I'll use the disabled blue badge and the sunflower lanyard examples—how do we modernize that, my experience with the blue badge, if I'm disabled, is all predicated on me being able to get access and fair access." Hence, as suggested, in future iterations, garnering trust and understanding within communities who already access the above schemes where credential attributes are commonplace, could assist in championing and advocating the benefits of using VC technologies. In short, the digitization of the previous schemes to assist vulnerable cohorts.

An expert embedded within marginalized communities further supported the assertion of "vulnerability and inclusion by design" advocated by Male1. Citing that "we work with young people, probably about six to twelve, to do the equivalent of a random control trial and saying, look, you know, this is what we want. This would be good to have our group as testers for such technology…you will have the support from people who want to be ambassadors or champions of what you are doing" (Male3). Furthermore, as this cohort utilizes social media for communication channels, the expert confirmed their "followers" would trust the communication and awareness would be generated across the most affected groups promoting engagement with VCs. Similarly, the element of trust was raised as pivotal to engendering uptake of new technologies to cohorts where the presumption is a lack of "tech savviness" (Female1). She continued, "we now talk to the majority of our members who do use this open banking, the majority have no problem with it whatsoever. You know, we are not talking on the marginal anymore, people are fine with it. I think because we are trusted with them (customers)."

From these excerpts, the difficulty perceived in relation to how to engage the target marginal and vulnerable groups, shows history and collaboration as a design mechanism to generate trust and adoption is feasible. Whether this causes complexity for the financial institutions who as we found are risk averse, especially since the global finance crisis, is beyond the scope of this project but raises areas of research to be explored around the concept of democracy and technology (Sclove, 2020).

*5.2.5. RQ3. How can financial firms use DID/VC technologies efficiently to improve support for vulnerable populations?*

Finally, we asked the experts to posit themselves as leading a financial institution and imagining how they could improve the support for vulnerable populations by their organization. What would be required based on their experience in introducing technologies.

Our expert (Male1) continued with the theme of vulnerability by design in advocating taking lessons learned from the development and implementation of open banking which culminated in application program interfaces (APIs) becoming understood and accepted technologies by financial providers (mainstream and alternative) and customers. He suggested that firms could "start with the principles of open banking, open, trusted custodian convener, independent, mutual, all of those things are not for profit. All those things that engender trust." Simply put, there is a route to implementation not too devoid of the recent open banking experience that applied to VC technology could assist financial firms understand how to better identify and support vulnerable customers. Furthermore, the "digital divide" must also be brought into consideration, the VCs solution is premised on technological access however, a workaround may be revealed in working closer with the communities in this cohort (Male1–3 and see van Dijk, 2019).

In addition, it was revealed that a mainstream provider had been in discussion with one expert in this regard. "A bank was asking that particular question about just sharing the passport number, nothing else. And by that passport number, it will be read and confirmed by the bank" (Male3). He suggests that the mainstream banks are aware that such technologies are under development and that this change is expected to come hence "very much aware of what's going on" based on this incident. Moreover, if financial providers support using technologies to reduce the stigma of "vulnerabilities" the use of "providing that sense of empowerment" is the key word to garner support from communities and for this question, will improve the mistrust of financial institutions to take the lead or collaborate in this space.

This expert considered if financial institutions can use the VC technologies to afford females a "voice and just…listen[ed] to them, and how they feel about their finances, they want to be, on top of things, generally, even if they have not got money…this will be welcomed by these women" (ibid.). It could be suggested that the concept of self-sovereign identity releasing attributes of credentials if offered by banks, as reported from this expert, would enable firms to "get this message across, the best way to do it is in the communities where these vulnerable people live and die" (Male1). Likewise, those dealing with vulnerable communities and providing responsible lending, the VC technology was viewed as an *"extension of the open banking"* innovation and no problems were envisaged in their network of similar providers improving their existing provision in supporting vulnerable customers (Female1). Moreover, all experts viewed the prototype as positive, including the title "Fairness for All" which was tackling the *"upstream problems"* to design better vulnerability solutions.

## 6. Policy Guidance in Balancing Innovation and Regulation

Based on the empirical work described in this report and the technology insights obtained from the implementation, we outline a set of recommendations for researchers and practitioners in the financial sector when considering the combination of DID and VC technologies, mostly focused on policy development and associated regulatory approaches.

- The tension between fulfilling regulatory requirements and pursuing innovation suggests that without involvement of the regulator, new technologies will find it difficult to gain deployment within the financial sector. From the interviews with experts, we conclude that without a leading role by the regulator (FCA in the UK) there is minimal chance of successful adoption of new technologies such as DID and VC.
- In terms of the importance of regulator support, the experts suggested that the industry can draw from the experience and practices of previous technological solutions to increase user uptake of these technologies. In particular, a comparison was made with Open Banking, which particularly in the UK has been strongly supported by regulatory developments, culminating in the 2017 Payment Services Regulations.
- Policy development and regulatory frameworks are not only relevant as a "stick" to enforce technological innovation. Regulation also plays an important role in developing societal trust in new technologies, both from the perspective of the value of new functionality and of the implications the deployment of new technologies may have on privacy and safety of consumers, in this case vulnerable consumers. Policy developers must therefore complement technological functionality with sufficient associated safeguards within any proposed policy and regulation.
- The disclosure of sensitive personal data such as vulnerabilities to financial service providers immediately creates challenges related to fulfilling GDPR requirements for the management and security of stored data. We believe that such concerns in part stem from uncertainty about the precise legalities of maintaining certain data, and should not be a show stopper. However, it puts a further onus on financial institution to provide for trained employees and sufficient support, also expanding the range of data under control of the data controller.
- On the other hand, the use of selective disclosure ensures data minimization, meaning that data that are not pertinent to a specific function or service are not collected unnecessarily. Assuming services are designed such that no information about the selected data is leaked from offered service or functionality, this reduces the privacy and security risks for financial service providers, protects consumers and facilitates compliance with existing regulation. We believe selective disclosure based on zero-knowledge proofs should become commonplace in future disclosure systems.
- Improvement of services for vulnerable consumers should be part and parcel of the business strategy of any financial service institution. Although it may be understandable that experts refer to the regulator to make possible solutions based on DIDs and VCs, they should be proactive in putting in place alternative initiatives to pragmatically improving the way vulnerable customers are served. As part of any Corporate Digital Responsibility strategy one would expect the industry to jointly progress the service quality for vulnerable customers, for instance through the use of a vulnerability registration service or solutions based on the blue badge or sunflower lanyard idea mentioned in Section 5.2.

The area of identification and credential disclosure may see considerable technological developments in upcoming years. In particular with respect to privacy preserving digital disclosure, systems can be expected to keep evolving in coming years to provide increased privacy guarantees while enabling new services. The principles of selective disclosure and self-sovereignty aim to support this objective and therefore deserve further attention and research.

The need to develop consumer trust in proposed technology played a critical role in our expert interviews. Although policy makers and regulators can support building that trust, technological developments should aim to achieve "trust by design." That is, systems should be designed such that privacy concerns are met technologically, but, importantly, are integrated in a manner that allow consumers to build trust in the technology. For instance, by making transparent to users the benefits of information disclosure, explaining how this information can enable better services and support in a privacy-preserving way.

## 7. Conclusion

This article explored the tension that exists within the financial services industry between the need to fulfil regulatory requirements and the desire to innovate. We studied this dilemma in the context of support for vulnerable customers, which in many jurisdictions is a regulatory requirement and in the UK is supported by guidance from the FCA. Implementing and improving support for vulnerable customers creates conflicts between the need to disclose sensitive personal information, a desire for privacy such as advocated through GDPR and key regulations that pertain to data disclosure (KYC) and management (GDPR).

We conclude from our research that to advance support for vulnerable customers all parties need to come together: financial services, citizen representatives as well as regulators and policy makers. Without regulatory support, it will be difficult to enable and gain acceptance for substantive technological innovation, because of the risk averse tendencies of the industry. At the same time, without innovative joined-up contributions from the industry around improved services, it will be impossible to shape a pathway toward more substantial technological innovation, for instance along the line of DIDs and VCs.

**Funding Statement.** This research was supported by a grant from the Bill and Melinda Gates Foundation (through The Alan Turing Institute) for "Finclusion"; the Knowledge Transfer Partnership between Atom Bank, Durham University and Newcastle University, funded by Innovate UK under grant Ref. KTP011059; and the UKRI EPSRC grant for "FinTrust", No. EP/R033595.

**Competing Interests.** The authors declare no competing interests exist.

**Author Contributions.** K.E. compiled and authored an earlier version of this article (as project deliverable for the Finclusion project); A.v.M. was main author of the submitted article with contributions from H.W. (technology survey), Z.M. and T.S. (policy considerations); S.F. and D.H. implemented and authored the description of the VC software; E.C. and P.E. were the main designers of the research effort based on the Knowledge Transfer Partnership with Atom Bank; K.E., K.C., M.N. and T.S. designed the qualitative study, while K.E. and T.S. conducted the interviews. All authors reviewed the submitted manuscript and K.E., T.S., H.W. and, A.v.M. prepared revisions.

**Data Availability Statement.** The code of the VC system is available from GitHub, for example, through the web site http://www.fintrustresearch.com. Reports Horsfall et al. (2021) and Elliott et al. (2021) are also available from the web site.

**Ethical Standards.** The research meets all ethical guidelines, including adherence to the legal requirements of the UK.

**Cite this article: Elliott K, Coopamootoo K, Curran E, Ezhilchelvan P, Finnigan S, Horsfall D, Ma Z, Ng M, Spiliotopoulos T, Wu H and van Moorsel A** (2022). Know Your Customer: Balancing innovation and regulation for financial inclusion. *Data & Policy,* 4: e34. doi:10.1017/dap.2022.23